# Localized Control of Curie Temperature in Perovskite Oxide Film by Capping-layer-induced Octahedral Distortion

S. Thomas[1], B. Kuiper[2], J. Hu[1,3], J. Smit[4], Z. Liao[2], Z. Zhong[5], G. Rijnders[2], A. Vailionis[4], R. Wu[1], G. Koster[2], J. Xia[1]

[1] Department of Physics and Astronomy, University of California, Irvine, Irvine, CA, 92697, USA
[2] MESA+ Institute for Nanotechnology, University of Twente, 7500AE Enschede, The Netherlands
[3] College of Physics, Optoelectronics and Energy, Soochow University, Suzhou, Jiangsu 215006, China
[4] Geballe Laboratory for Advanced Materials, Stanford University, Stanford, California 94305, USA
[5] TU Wien, Vienna University of Technology, Karlsplatz 13, 1040 Wien, Austria

With reduced dimensionality, it is often easier to modify the properties of ultra-thin films than their bulk counterparts. Strain engineering, usually achieved by choosing appropriate substrates, has been proven effective in controlling the properties of perovskite oxide films. An emerging alternative route for developing new multifunctional perovskite is by modification of the oxygen octahedral structure. Here we report the control of structural oxygen octahedral rotation in ultra-thin perovskite $SrRuO_3$ films by the deposition of a $SrTiO_3$ capping layer, which can be lithographically patterned to achieve local control. Using a scanning Sagnac magnetic microscope, we show increase in the Curie temperature of $SrRuO_3$ due to the suppression octahedral rotations revealed by the synchrotron x-ray diffraction. This capping-layer-based technique may open new possibilities for developing functional oxide materials.

Email: xia.jing@uci.edu

Ultra-thin films offer unique possibilities for fabricating novel optical [1], electronic [2] and spintronic [3] devices. In particular, functional complex oxide heterostructures [4] have recently attracted wide attention, in part because the properties of these oxides can be tailored via mechanical strain through the choice of substrates, allowing one to create new materials with desired properties. Large enhancement of ferroelectricity by depositing on strained substrates has been demonstrated in thin films of $SrTiO_3$ (STO) [5] and $BaTiO_3$ [6]. Tuning of ferromagnetism with "strain engineering" has been realized in $SrRuO_3$ (SRO) [7]. An emerging alternative to strain engineering is to modify the oxygen octahedral structure [8]. It has been theorized that distortions to the oxygen octahedra could have large impacts on oxide film's electrical and magnetic properties [9]. Such distortions have recently been observed to extend several molecule layers (ML) into SRO film [10].

Here we explore a different method of controlling the octahedral distortion, by growing a STO capping layer on top of the SRO film with the hope of tuning SRO's properties. Since the lateral strain in SRO is largely determined by the substrate, this approach should offer independent controls of strain and octahedral distortion. Experimentally we indeed observe a large enhancement of Curie temperature (Tc) in SRO with just a few ML of STO capping. Using synchrotron x-ray diffraction, we show that the STO capping changes the oxygen octahedral rotation in SRO film without altering its lateral strain. Density functional theory (DFT) calculations confirm that this octahedral distortion is the cause for the observed Tc enhancement. Using a scanning Sagnac microscope, we demonstrate localized control of Tc in SRO with spatially patterned capping layer: a unique capability that is lacking in substrate-based approaches. These results may point to new opportunities for functional complex oxides.

The perovskite oxide $SrRuO_3$ (SRO) used in this study [11] is a rare case of 4d itinerant ferromagnet with near-perpendicular anisotropy. It is an ideal electrode material to incorporate with a variety of functional complex oxides, including high temperature superconductors. And its itinerant ferromagnetism makes it a possibility as a spin current injector [12,13] or as a spin memory [14]. SRO samples used in this work were grown by pulsed laser deposition (PLD), with film thickness determined by reflection high-energy electron diffraction (RHEED) oscillations. The AFM images of the films (Fig.1b) showed single ML steps indicating atomically smooth surface typical for a step-flow like growth mode pointing to a single crystalline quality thin film. These films were grown on $DyScO_3$ (DSO) substrates, which induce a tensile strain in the SRO films. STO capping-layers of varying thicknesses were grown on top of the SRO film *in situ*.

To measure and spatially image the magnetization in ultra-thin SRO films we employ a scanning microscope version of the loop-less Sagnac interferometer [15], which measures the polar magneto-optic Kerr effect (MOKE) by interfering circularly polarized lights of opposite chiralites. Thus it rejects artifact signals that are usually time-reversal symmetry-invariant [16], and has achieved nano-radian level Kerr sensitivity [15]. Since the SRO films used in this experiment are much thinner than the optical skin depth, the measured polar Kerr signal is directly proportional to the film's perpendicular magnetization component [17].

The polar Kerr signals from uncapped SRO films are found to be consistent with a previous study [18]. The ferromagnetic transition (Curie) temperature Tc can be determined precisely by monitoring the remnant Kerr signal at zero magnetic field while warming up the sample. The blue curves in Fig.1c and d represent such measurements for a 10-ML and a 20-ML thick SRO films exhibiting the Tc values of 131 K and 136 K respectively, that are close to SRO films grown on STO substrate [18].

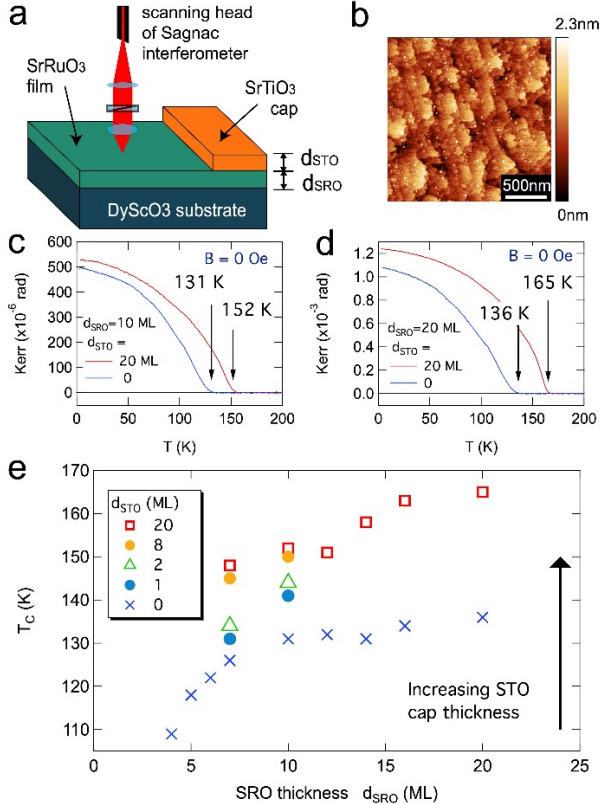

FIG. 1. (Color online) $T_C$ enhancement in SRO by STO capping layer. **(a)** Sagnac Kerr measurement configuration. **(b)** AFM image of 10ML SRO on DSO substrate with 8ML STO capping-layer. **(c)** Tc comparison between two 10ML SRO films: with a 20ML STO capping-layer; uncapped. **(d)** Tc comparison between capped and un-capped 20ML SRO films. **(e)** $T_C$ of SRO films for several combinations of SRO film thickness $d_{SRO}$ and STO capping thickness $d_{STO}$. STO capping enhances Tc in all samples.

We find that adding a few-ML thick STO capping layer induces a dramatic enhancement in Tc. As shown in Fig.1c, a 10 ML SRO film with a 20 ML STO capping layer (red curve) has a Tc of 152 K, 21 K higher than its uncapped counterpart (blue curve). A 20 ML thick SRO film with a 20 ML STO capping shows an even larger Tc enhancement of 29 K (Fig.1d). Samples with different combinations of SRO film thickness and STO capping thickness were grown and measured. And the results are summarized in Fig. 1e, where the vertical axis is Tc and the horizontal axis is SRO film thickness $d_{SRO}$. Different markers represent different STO capping thickness $d_{STO}$. Tc enhancements up to 30 K are observed in all these combinations. We note that Tc in some configurations has even surpassed the 155 K value in bulk SRO [11].

To understand the origin of the observed Tc enhancement, we performed detailed studies of the changes in SRO thin film structure induced by the STO capping layer employing laboratory and synchrotron x-ray diffraction (XRD). Two samples were examined with the XRD: an uncapped 7 ML thick SRO film (DSO31) and a 7 ML thick SRO film capped with 8 ML STO (DSO84). Both samples were epitaxially grown on DSO(110) substrate in the same growth batch. Reciprocal space maps (RSM) of uncapped and capped SRO films were collected around orthorhombic/pseudocubic (420)o/(103)p, (240)o/(-103)p, (332)o/(013)p and (33-2)o/(0-13)p Bragg peaks using X'Pert materials research diffractometer at the Stanford Nano Shared Facilities. The RSMs (see supplementary Fig. S1) revealed that SRO layers and the STO cap were coherently strained to the single crystal DSO substrate along [1-10] and [001] in-plane directions. For uncapped sample the pseudocubic unit cell of SRO film was determined to be monoclinic and tilted only along in-plane orthorhombic [1-10] direction of the DSO substrate. No tilt was observed along [001] direction confirming monodomain growth of monoclinic SRO layer. To quantify the out-of-plane lattice parameter and tilt angle of a pseudocubic SRO unit cell, subsequent L-scans around pseudocubic (103), (-103) and (013) Bragg peaks were collected at the beam line 7-2 of the Stanford-Synchrotron-Radiation-Light-Source (SSRL). The L-scans around (103), (-103) Bragg peaks and corresponding peak fits are shown in Fig. 2. The accurate peak positions of the SRO layer and the STO cap were obtained by fitting the experimental XRD profiles with the corresponding DSO, SRO and STO Bragg peaks assuming Pearson VII peak shapes. Previous x-ray diffraction studies established that thicker SRO films grown on DSO(110) substrates under tensile stress exhibit tetragonal unit cell with the pseudocubic unit cell parameters $a_p \neq b_p \neq c_p$ and the tilt angle $\beta_p = 90.00°$ [19], following the notations of axes and tilting angles in Supplementary Material Fig. S2. Note that $\beta_p$ angle is used to describe the tilt of the pseudocubic unit cell away from the [001] out-of-plane direction. Such unit cell exhibits $a^+b^-c^0$ rotational pattern with suppressed out-of-plane rotations due to tensile strain and finite rotations along perpendicular in-plane directions [20]. RuO$_6$ octahedral rotation angles α($a^+$), β($b^-$), γ($c^-$) according to Glazer notation signify the in-phase, out-of-phase rotations and the tilt magnitudes. In Fig. 2 we show that, unlike in the thicker SRO films, in ultrathin SRO layers on DSO substrate the out-of-plane rotations to some degree are preserved. In Glazer notation for pseudocubic unit cell with $a^+b^-c^0$ rotational pattern, $\beta_p = 90°$. The pseudocubic unit cell with $a^+b^-c^-$ rotations will have a monoclinic shape where

$\beta_p$ deviates from 90° and the crystallographic unit cell will be orthorhombic in this case.

The pseudocubic lattice parameters of all the layers obtained by XRD analysis are summarized in Fig. 2e. The SRO lateral lattice parameters $a_p$ = 3.952 Å and $b_p$ = 3.957 Å are identical between DSO31 and DSO84 samples, indicating that the STO capping doesn't change the lateral strain in SRO, unlike in previous substrate-based strain experiments [5-7,21]. The pseudocubic SRO unit cell in DSO31 sample is monoclinic with $\beta_p$ = 89.67°, indicating that SRO film possesses $a^+b^-c^-$ rotational pattern with nonzero out-of-plane rotations. The pseudocubic tilt angle $\beta_p$ of the SRO layer is slightly larger than in bulk SRO which can be attributed to the effect of tensile strain induced by the DSO substrate. In contrast to the uncapped SRO film in DSO31 sample, the STO-capped sample DSO84 exhibits tetragonal unit cell with pseudocubic tilt angle $\beta_p$ = 89.98°, resulting in a suppressed out-of-plane octahedral rotations. The pseudocubic unit cell with $\beta_p \approx 90°$ is a direct consequence of unique strain accommodation by the SRO layer in the presence of the STO cap.

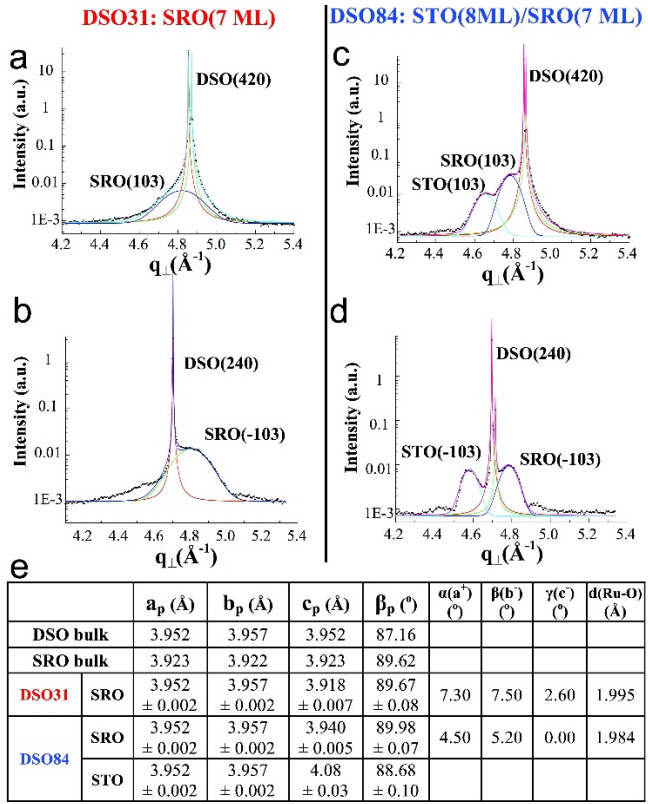

FIG. 2. (Color online) Structural changes. XRD L-scans around $(420)_o/(103)_p$ and $(240)_o/(-103)_p$ Bragg reflections and corresponding peak fittings: **(a)**, **(b)** for uncapped sample DSO31, and **(c)**, **(d)** for STO capped sample DSO84. **(e)** Pseudocubic unit cell parameters and RuO$_6$ octahedral rotation angles of SRO and STO layers.

Density functional theory (DFT) calculations were performed for SRO with the experimental lattice structure in either the DSO31 or DSO84 sample. Indeed, the later has an exchange energy (defined as $\Delta E_{ex} = E_{AFM} - E_{FM}$) of 39.0 meV, larger than that of former, 37.1 meV. This indicates that the structural deformation in the SRO layer is the main cause for the capping-induced increase of the Curie temperature Tc. To further disentangle the two factors: lattice strain and the rotation of corner-connected oxygen octahedra, we investigated the magnetic properties of SRO in both orthorhombic and tetragonal structures as depicted in the insets in Fig. 3a. The ground state orthorhombic phase of SRO has an octahedral tilt angle $\theta = \cos^{-1}(\cos\alpha(a^+)\cos\beta(b^-))$ of 9.5° away from the $c$ axis, and is 0.3 eV/f.u. lower in energy compared to the metastable tetragonal phase. In both phases, $\Delta E_{ex}$ appears to respond rather slowly to the hydrostatic strain $V/V_0$, as shown in Fig. 3a, agreeing with early experiments [22,23], indicating that the lattice strain cannot produce significant change of Tc of SRO. In contrast, $\Delta E_{ex}$ of the tetragonal phase is much higher than that of the orthorhombic phase. Moreover, $\Delta E_{ex}$ increases very steeply with the reduction of rotational angle away from the orthorhombic phase as shown in Fig. 3b. As shown in the Supplementary Material, our DFT calculations for a STO/SRO/STO film indicate that the octahedral tilts of SRO are reduced by 3.3° to 0.5° from the interfacial to the third SRO layers. This angle is also reduced by about 0.28° in our calculations for bulk structures from that mimics the DSO31 sample to that mimics the DSO84 sample. Therefore, we may attribute the dramatic enhancement of $T_C$ in Fig. 1 to the reduction of octahedral rotation/tilting caused by STO-capping. Furthermore, the exchange interaction near the interface is stronger than in the interior region. Nevertheless, SRO is an itinerant magnet, the itinerant electrons spread across the layers, we thus have a singular Tc value even though different layers have different octahedral angles.

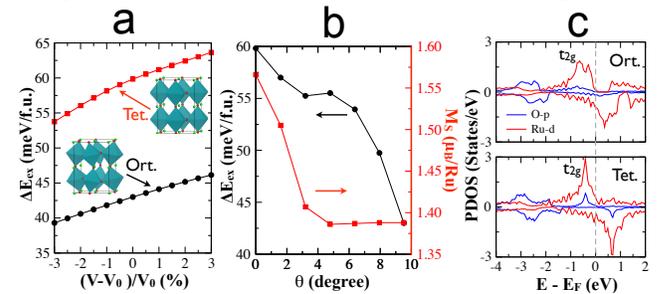

FIG. 3 (Color online) DFT calculation results. **(a)** $\Delta E_{ex}$ of orthorhombic and tetragonal SRO as a function of hydrostatic strain. The insets show the atomic structures of orthorhombic and tetragonal SRO. The red and green spheres stand for the oxygen and strontium atoms, respectively. The ruthenium atoms locate at the centers of the pseudo-octahedra. **(b)** $\Delta E_{ex}$ and magnetic moment Ms as functions of octahedral tilt angle θ. **(c)** Projected density of states (PDOS) on O and Ru atoms for the strain-free SRO.

The curves of the projected density of states (PDOS) in Fig. 3c indicate that the tetragonal SRO becomes metallic in both spin channels, in contrast to the half-metallic feature of the orthorhombic SRO. The spin magnetic moment ($M_S$) of the orthorhombic SRO is 2.0 $\mu_B$ per SRO formula unit, 1.39 $\mu_B$ on Ru, in good agreement with previous calculations and experimental measurements [24,25]. $M_s$ of the tetragonal SRO is 2.24 $\mu_B$ per SRO formula unit (1.57 $\mu_B$ on Ru), which can be interpreted as the result of higher Stoner instability of SRO in the tetragonal phase. The orthorhombic SRO is an itinerant ferromagnet with the delocalized $t_{2g}$ states solely in the minority spin channel mediating the exchange interaction among Ru atoms. The availability of the itinerant electronic states in both spin channels near the Fermi level leads to a stronger exchange interaction between Ru atoms and hence higher Tc. The electronic and magnetic properties of both the orthorhombic and tetragonal SRO do not change much under certain external strain (Supplemental Fig. S4).

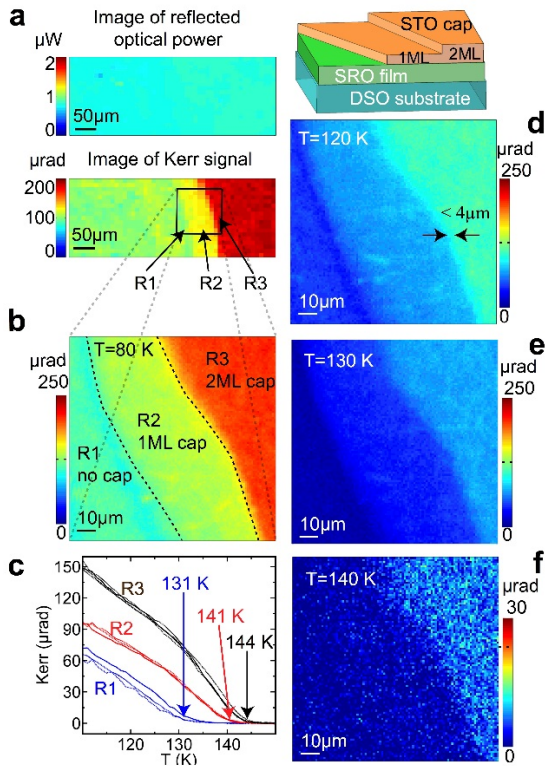

FIG.4 (Color online) Local Tc control by patterning the STO capping layer. **(a)** Scans of the reflected optical power (top) and Kerr signal (bottom) over the transition area between etched and unetched regions of a STO-capped 10ML SRO film. Regions R1, R2 and R3 correspond to 0, 1, and 2 ML of STO capping. **(b), (d), (e), (f)** Fine scans of the transition region at 80 K, 120 K, 130 K and 140 K respectively. **(c)** Temperature dependent remnant Kerr signal at different spots showing quantitatively uniform Tc within each region, but different Tc between regions.

Finally, the reported capping-based technique allows us to control Tc locally. Unlike bulk substrates, the STO capping layer can be patterned lithographically. As shown in Fig. 4, we pattern the 2-ML-thick STO capping layer of a 10-ML SRO film using standard photo-lithography and etchant mixture of HF, $HNO_3$, and $H_2O$ [7] that selectively etches STO. The result is a structure illustrated in the inset of Fig. 4d: region R3 is covered with 2 ML STO, region R1 has no STO capping, and region R2 is covered with 1 ML STO. Since ML-thick STO capping barely changes the optical property, we see uniform optical reflectivity across all three regions. In contrast, all three regions are clearly visible in the Kerr image (Fig. 4a), which were taken at zero magnetic field and hence the color-bared Kerr signals represent remnant magnetization. Note that the optical cryostat used for scanning imaging in Fig. 4 is equipped with a smaller magnet insufficient to fully saturate SRO's magnetization at low temperatures. Consequently, the remnant magnetization is smaller than what was shown in Fig. 1 where a non-imaging cryostat with a bigger magnet is used. We show in Figs. 4b, d–f the Kerr images at different temperatures across ferromagnetic transitions. Consistent with the findings summarized in Fig. 1e, ferromagnetism onsets at very different Tc for regions R1, R2 and R3. To precisely determine Tc in each region, we perform temperature dependent measurements on several spots as shown in Fig. 4c. While the Curie temperature is uniform in each region, it is quite different between regions: 134 K for R1, 141 K for R2, and 144 K for R3 as a results of different capping layer thickness. The spatial sharpness of this local control is at least as good as our optical spatial resolution of 2 μm (Fig. 4d), and it is likely to be ultimately determined by the sharpness of the edge of the STO capping. Furthermore, the restoration of $T_c$ in the region R1 serves as an evidence that the diffusion of Ti atoms into SRO is not responsible for the capping-induced enhancement of $T_c$ in this work, as they should not be completely removed by etching.

In conclusion, an enhancement of $T_C$ in SRO thin-films due to a STO capping has been demonstrated. XRD measurements reveal diminished oxygen octahedral tilt away from the *c*-axis in the capped SRO layers. This leads to enhanced magnetic exchange energy and hence Tc enhancement according to DFT calculations. This capping-layer-based approach allowed us to perform local Tc tuning, and may provide a new route for property engineering of ultra-thin complex oxide materials.

This work is supported by NSF award DMR-1350122. Samples growth is supported by Netherlands foundation for Scientific Research (NWO). J. Hu and R. Wu acknowledge support from DOE-BES grant No. DE- FG02-05ER46237 and computing time allocation from NERSC. Part of this work was performed at the Stanford Nano Shared Facilities (SNSF), supported by the NSF award ECCS-1542152.

**Supplementary Materials for:**

# Localized Control of Curie Temperature in Perovskite Oxide Film by Capping-layer-induced Octahedral Distortion

S. Thomas[1], B. Kuiper[2], J. Hu[1,3], J. Smit[4], Z. Liao[2], Z. Zhong[5], G. Rijnders[2], A. Vailionis[4], R. Wu[1], G. Koster[2], J. Xia[1]

[1] Department of Physics and Astronomy, University of California, Irvine, Irvine, CA, 92697, USA

[2] MESA+ Institute for Nanotechnology, University of Twente, 7500AE Enschede, The Netherlands

[3] College of Physics, Optoelectronics and Energy, Soochow University, Suzhou, Jiangsu 215006, China

[4] Geballe Laboratory for Advanced Materials, Stanford University, Stanford, California 94305, USA

[5] TU Wien, Vienna University of Technology, Karlsplatz 13, 1040 Wien, Austria

## Materials and Methods:

## Sample Growth:

SRO samples used in our experiment were grown by Pulsed Laser Deposition (PLD). The samples were grown in a vacuum chamber with a background pressure of $10^{-7}$ Torr. A 248 nm wavelength KrF excimer laser was employed with typical pulse lengths of 20–30 ns. The energy density on the target is kept at approximately 2.1 J/cm$^2$. Films were deposited with a laser repetition rate of 4 Hertz, with the substrate temperature at 700 C. Films were grown on ScO terminated DSO substrates with [110] orientation or TiO$_2$ terminated STO surfaces with [110] orientation [1]. DSO has a lattice roughly .574% larger than SrRuO3, whereas STO has a lattice that is roughly .446% smaller. The thickness of the films ranges from 3–35 monolayers (ML), as determined by calibrating on thicker samples with x-ray reflectivity. The reproducibility of the thickness was found to be very high.

## XRD Measurement Details:

### *XRD Measurement:*

To investigate the change in structural properties of SRO thin films induced by the capping layer, two samples were selected for x-ray diffraction (XRD) analysis: uncapped 7 u.c. thick SRO layer and 7 u.c. thick SRO layer capped with 8 ML of STO layer. Both samples were coherently grown on DSO(110) substrates. XRD measurements were performed using both laboratory and synchrotron sources. Reciprocal space maps (RSM) of both samples were collected around orthorhombic/pseudocubic $(420)_o/(103)_p$, $(240)_o/(-103)_p$, $(332)_o/(013)_p$ and $(33\text{-}2)_o/(0\text{-}13)_p$ Bragg peaks using X'Pert materials research diffractometer at the Stanford Nano Shared Facilities (SNSF). In order to quantify the out-of-plane lattice parameter and tilt angle of a pseudocubic SRO unit cell, subsequent L-scans around pseudocubic (103) and (-103) and (013) Bragg peaks were collected at the beam line 7-2 of the Stanford Synchrotron Radiation Lightsource (SSRL).

Reciprocal space maps (RSM) revealed that SRO layers and the STO cap were coherently strained to the single crystal DSO(110) substrate along [1-10] and [001] in-plane directions. The pseudocubic unit cell of the SRO film was determined to be monoclinic and tilted only along in-plane orthorhombic [1-10] direction of the DSO substrate. No tilt was observed along in-plane [001] direction confirming monodomain growth of the monoclinic SRO layer.

*Reciprocal Space Maps:*

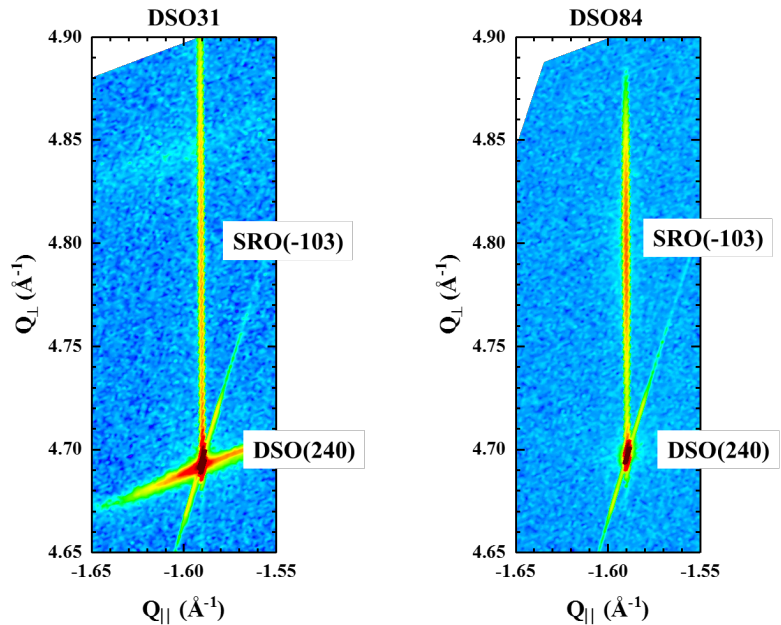

**Fig. S1.** Reciprocal space maps (RSM) around DSO(240) Bragg reflection for DSO31 and DSO84 samples. The maps show that SRO thin films in both samples are coherently strained to the DSO substrate. In order to make visible the SRO(-103) Bragg peak, the intensity of the DSO substrate was made intentionally saturated.

*Fitting XRD profile to obtain lattice parameters:*

The L-scans around (103), (-103) Bragg peaks and corresponding peak fits are shown in the main text Fig. 2 a-d. The accurate peak positions of the SRO layers and the STO cap were obtained by fitting the experimental XRD profiles with the corresponding DSO, SRO and STO Bragg peaks assuming Pearson VII peak shapes. The pseudocubic lattice parameters of SRO layers and STO cap were obtained from the fitting results and presented together with the bulk DSO and SRO data in the main text Fig. 2 e.

The monoclinic tilt angle of the pseudocubic unit cell was determined from the separation of (103) and (-103) Bragg peaks of the SRO layers and the STO cap.

*Relationship between the crystallographic axes, lattice parameters and angles of the pseudocubic unit cell:*

Figure S2 demonstrates the relationship between the lattice parameters and the angles with respect to the crystallographic directions of the orthorhombic and pseudocubic unit cell. For the orthorhombic unit cell notation the out-of-plane axis is along $[110]_o$ direction, while for the pseudocubic unit cell notation the $[001]_p$ direction is defined as the out-of-plane direction, while $[100]_p$ and $[010]_p$ directions are in the plane of the film. The $\alpha(a^+)$, $\beta(b^-)$, $\gamma(c^-)$ angles discussed in the main text are the octahedral rotation angles according to Glazer notation for pseudocubic unit cell, where the $a^+$, $b^-$ and $c^-$ are the indices signifying in-phase, out-of-phase rotations and the tilt magnitudes [2].

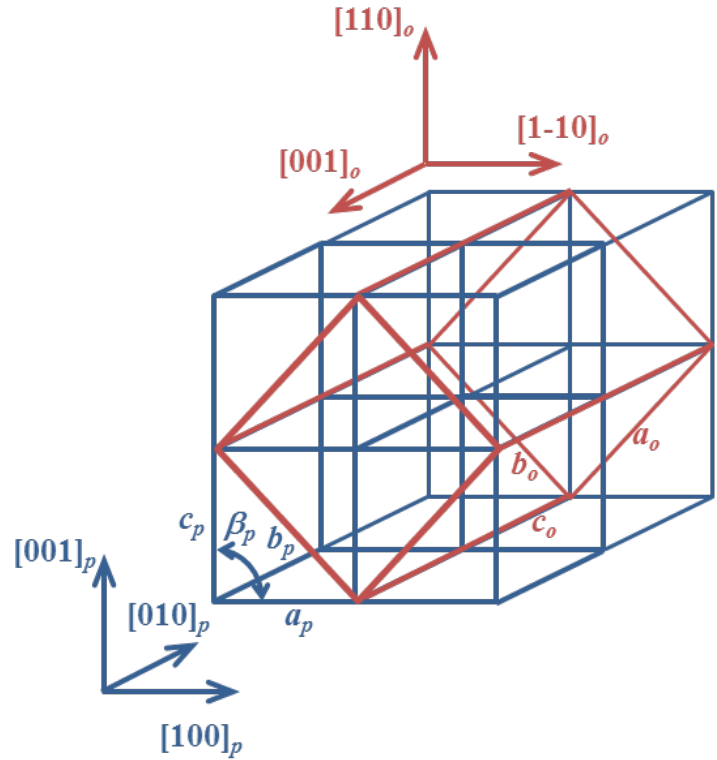

**Fig. S2.** Schematic view of the relationship between the crystallographic axes, lattice parameters and angles of the pseudocubic unit cell and orthorhombic unit cell. The $[001]_p$ direction is defined as the out-of-plane direction, while $[100]_p$ and $[010]_p$ directions are in-plane.

## DFT Computational Details:

The density functional calculations were performed with the Vienna ab-initio simulation package (VASP) [3,4]. The interaction between valence electrons and ionic cores was described within the framework of the projector augmented wave (PAW) method [5,6]. The spin-polarized generalized-gradient approximation (GGA) was used for the exchange-correlation functional, since it is better for the description of SRO [7] than local density approximation (LDA) with and without the Hubbard U correction [8]. The energy cutoff for the plane wave basis expansion was set to 400 eV. The Brillouin zone was sampled by a dense k-point mesh 15×15×11. The atomic positions were fully relaxed using the conjugated gradient method for the energy minimization procedure, with a criterion that requires the force on each atom smaller than 0.01 eV/Å.

### *Octahedral rotational angle, definition and results for model systems:*

In $SrRuO_3$, the octahedra distortion (deviation of O-Ru-O angle from 90°) is much smaller than the octahedra rotation. The former is about ±1.5° for most cases, and it does not change much when the hydrostatic strain is applied. The octahedra tilt angles are as large as 10°, and the octahedra rotate not only about the $c$ axis but also about the $a$ and $b$ axes, as shown in the Fig.S3 bellow. The horizontal axes of Fig. 3b is the $\theta$ angle as defined in Fig. S3 below.

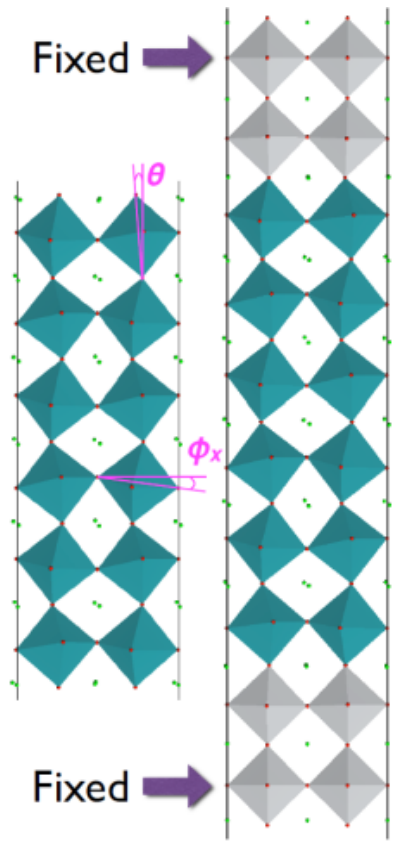

**Fig. S3.** Schematic view of the octahedral tilt angles of the SRO and STO/SRO films. For the calculations of the STO/SRO films, the octahedra in the topmost STO are fixed with zero rotation.

|  | 1st (interfacial) | 2nd | 3rd (central) |
|---|---|---|---|
| | | SRO(6L) | |
| $\theta$ | 7.10 | 9.52 | 10.34 |
| $\phi_x$ | 10.53 | 10.28 | 10.19 |
| $\phi_y$ | 10.50 | 10.65 | 10.77 |
| | | STO(2L)/SRO(6L)/STO(2L) | |
| $\theta$ | 3.83 | 8.78 | 9.83 |
| $\phi_x$ | 10.49 | 10.37 | 10.20 |
| $\phi_y$ | 10.61 | 10.79 | 10.79 |

**Table S1:** the calculated octahedral tilt angles for the two models in Fig. S3. While the topmost STO layer was fixed in their bulk structure (zero rotation), the value of $\theta$ for the interfacial STO layer is 1.04 degree.

### Conversion between octahedral rotation angles (α, β, γ) used in crystallography and octahedral tilt angles (θ, $\phi_x$, $\phi_y$) used in DFT calculation:

In this paper, octahedral rotation angles α(a$^+$), β(b$^-$), γ(c$^-$) are obtained from XRD measurements following the convention in crystallography. Octahedral tilt angles θ, $\phi_x$, $\phi_y$ (Fig. S3) are used in DFT calculations since they describe the rotation of chemical bonds in a clearer fashion. In the table in Figure 2e we show octahedral rotation angles that were obtained from pseudocubic SRO unit cell lattice parameters. One can see that the tilt of a monoclinic unit cell by ~0.3° induces rather significant Ru-O-Ru bond angle changes. The DFT calculations use the octahedral tilt angles θ, $\phi_x$, $\phi_y$ with respect to the $a_p$, $b_p$, and $c_p$ pseudocubic lattice parameters.

The relation between octahedral tilt angles and θ, $\phi_x$, $\phi_y$ angles used in the DFT calculations are as follows:

$$\theta = \cos^{-1}(\cos\alpha(a^+) \, \cos\beta(b^-)),$$
$$\phi_x = \cos^{-1}(\cos\beta(b^-) \, \cos\gamma(c^-)),$$
$$\phi_y = \cos^{-1}(\cos\gamma(c^-) \, \cos\alpha(a^+)),$$

,where α(a$^+$), β(b$^-$), γ(c$^-$) are octahedral rotation angles around $a_p$, $b_p$, and $c_p$ pseudocubic unit cell axes. And θ, $\phi_x$, $\phi_y$ are octahedral tilt angles as illustrated in Fig. S3.

### Evolution of Density of State of SRO with lattice size.

Figure S4 below shows that, without the change or octahedral rotation angle, the orthorhombic (tetrahedral) SRO remain to be half-metallic (metallic) within the range of hydrostatic strain from -3% to 3%.

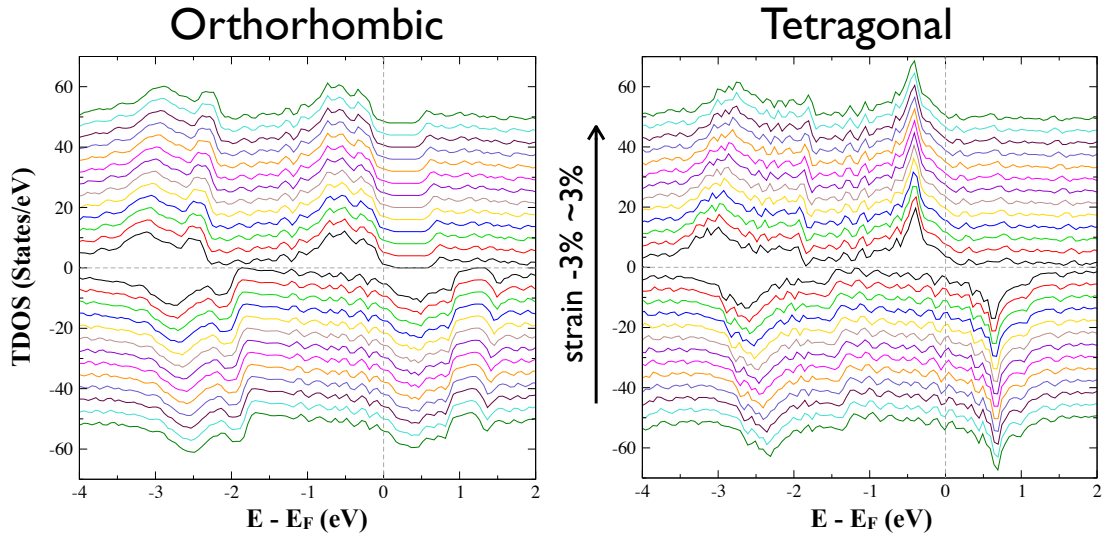

**Fig. S4.** *DFT Calculated Density of States. Total density of states (TDOS) of orthorhombic and tetragonal SRO under hydrostatic strain from -3% to 3%.*